\input{psfig.sty}

\documentclass[12pt,emulateapj5]{aastex}
\shorttitle{??}
\shortauthors{Peres et al.}
\usepackage{amsmath}            

\begin{document}

\title{Are coronae of late type stars made of solar-like structures?
The F$_X$-HR diagram and the pressure-temperature correlation.
}
\author{ G. Peres}
\affil{Dip. di Scienze Fisiche ed Astronomiche, Sezione di Astronomia, \\
	Piazza del Parlamento 1, 90134 Palermo, Italy -- 
	peres@oapa.astropa.unipa.it} 
\author{S. Orlando} 
\affil{INAF - Osservatorio Astronomico "G.S. Vaiana",\\ 
	Piazza del Parlamento 1, 90134 Palermo, 
	Italy -- orlando@oapa.astropa.unipa.it} 
\author{F. Reale} 
\affil{Dip. di Scienze Fisiche ed Astronomiche, Sezione di Astronomia, \\
	Piazza del Parlamento 1, 90134 Palermo, Italy -- 
	reale@oapa.astropa.unipa.it}
 
\begin{abstract}

This work is dedicated to the solar-stellar connection, i.e. the close
similarity of the Sun and the late-type stars; in particular this work
shows that stellar coronae can be composed of {\em X-ray emitting}
structures similar to those present in the solar corona.  To this end
we use a large set of ROSAT/PSPC observations of late-type-stars of all
spectral types and activity levels, and a large set of solar X-ray data
collected with Yohkoh/SXT.

Solar data have been analyzed and formatted to study the Sun as an
X-ray star; they include observations of the solar corona at various
phases of the cycle and data on various kinds of {\em X-ray} coronal
structures, from flares to the background corona, i.e. the most quiet
regions.

We use the X-ray surface flux ($F_X$) vs. spectral hardness ratio (HR)
diagram as a fundamental tool for our study.  We find that $F_X$ is
strongly correlated to HR in stellar coronae, in the solar corona at
all phases of the cycle, and in the individual solar coronal
structures; all the above follow the same law.  Schmitt found the same
correlation in stellar coronae.

We therefore claim that coronae of late type stars are formed with
{\em X-ray} structures very similar to the solar ones, since their
behavior is identical to that of the solar coronal structures and of
the whole solar corona. The spatial location of the {\em X-ray}
structures on the star, however, can be very different than on the
Sun.

In this scenario, the fraction of the stellar surface covered with
active regions and with their bright cores increases with activity; the
most active stars are brighter and hotter than if they were entirely
covered with active regions so they can be explained only with the
additional presence of several flares (or flare-like structures) at any
time.

On the basis of the $F_X$ vs. HR correlation, {\em corresponding to
$F_X \propto T^6$}, we then derive a set of new laws relating the
temperature, pressure, volumetric heating and characteristic loop
length of the coronal plasma, on all the late type stars. Also,
individual solar coronal structures and the whole solar corona, follow
the same laws. These laws also agree with recent findings of higher
plasma density at higher temperatures in stellar coronae.

We claim that the strong correlation between surface flux and
temperature and the laws mentioned above are just the effect of more
fundamental physical mechanisms driving the coronal structures of all
the late-type stars from the emergence of new magnetic structures to
their dispersal and dissipation.

\end{abstract}

\section{Introduction}

Since the seminal work by Vaiana et al. (1981) it has become clear
that all late type stars share the same basic coronal characteristics:
the presence of thermal plasma at several million degrees, its magnetic
confinement, the presence of flares etc.  Also the basic plasma physics
governing coronae is probably very similar in all these stars.  On the
other hand, coronae cover a large range of X-ray luminosity (even among
stars of the same spectral type and luminosity class) and a few scientists
think that the coronae of the most active late-type stars are different
from the solar corona, albeit the crucial differences have rarely been
stated clearly.

Two otherwise identical stars may have rather different levels of stellar
activity, i.e. X-ray luminosity, spottedness, UV luminosity etc., because
of very different rotation speed (Pallavicini et al., 1981) and age
(for a review on the subject Micela, 2002); the
speed probably determines how effectively the stellar magnetic dynamo generates
magnetic fields.

One may then expect that many differences among late-type stars coronae
are just due to rather different regimes of the dynamo at work inside
stars, while the basic plasma phenomena occurring in the outer magnetized
coronal plasma are probably the same and only the amount and scale of such
phenomena are different.  Along this line of thought, this work explores
to which extent most of the differences found among the various coronae
can be attributed to their different composition in terms of various
kinds of coronal structures (ranging from the relatively faint and
cool structures of the background corona to the very bright and hot flaring
regions) and to the amount of {\em X-ray emitting} coronal structures present; on the other
hand we explore to which extent the plasma phenomena and the structures
involved in the coronae are similar even in stars of very different
activity levels.  According to this paradigm, the solar corona offers
the opportunity to study and understand better the building blocks of
any late-type-star corona.  To this end we consider the average surface
flux ($F_X$) and the spectral hardness ratio (HR) in the ROSAT/PSPC band,
of late-type stars with those of solar coronal structures; we also study
the strong $F_X$--HR correlation (Schmitt, 1997; henceforth S97).

In the following we do not claim that all the aspects of solar corona
are the same in the other stars but that the fundamental features are
the same; for instance there is evidence of polar active regions and of
large polar flares (e.g. Schmitt and Favata, 1999; Brickhouse, Dupree
and Young, 2001; {\em see also Schrijver and Title, 2001). We will
concentrate on the X-ray emission from stellar coronae and we will
compare the solar and the various stellar cases and through analogies
and differences we will infer (or just constraint) features of X-ray
emitting structures present in stellar coronae.

A few works along this line of thought have been published, either
studying the temperature vs. emission measure per unit area on active
stars (Schrijver et al. 1984; Jordan and Montesinos, 1991) or correlating
activity parameters (e.g. luminosity in given bands or emission measure)
with temperature (Schmitt et al., 1990; Guedel et al.,
1997; Preibish, 1997). More in detail, Schrijver et al. (1984) analyzed
a sample of 34 late-type stars observed with the Einstein telescope,
related the relevant coronal temperature and emission measure per unit
area and rotation rate, determining scaling laws among them and relating
the first two through loop models. Jordan and Montesinos (1991) studied
various scalings among emission measure, coronal temperature, surface
gravity and Rossby number. Schmitt et al. (1990) performed an extensive
survey of late-type stars observed with the Einstein telescope, studying
and relating the coronal features and showing some differences among M
stars, F and G stars, giants and RS CVn stars. 
Guedel et al. (1997) studied the "X-ray Sun in time"
with the help of various X-ray observations of solar-like stars of
different ages; they found a power law dependence of the total X-ray
luminosity with the temperature of the hottest component of a
two-temperature spectral fit. Preibish (1997) studied a sample of young
late-type stars and found a power law dependence of total X-ray
luminosity vs. the temperature of the hottest component.}

This work is based on methods and results developed in our program to
study the solar-stellar connection in X-rays, mostly based on ROSAT/PSPC
and Yohkoh/SXT data.  {\em Our approach allows us to tie directly, with
just little elaboration and interpretation, the vast amount of data
collected with the two aforementioned observatories. Also our method and
scope is novel with respect to previous works, since we use 
solar data (made homogeneous to stellar data) as a template and a guide.
Our findings are eventually interpreted in
terms of coronal loops scaling laws.} ROSAT has provided the largest {\em homogeneous} data set of
stellar coronal observations so far collected;  this data set is an
important reference for studies of stellar coronae because:  i) the
relevant data have been collected with just one instrument, ii)
instrumental and observing conditions were relatively stable, iii) a
large number of stars of different spectral types and luminosity
classes have been observed and iv) the data set has an extended time
coverage. For analogous reasons, Yohkoh/SXT data are {\em equally} important to study
the solar corona.  It is, therefore, natural to refer both to ROSAT and
to Yohkoh data in order to model a scenario of coronae on late-type
stars and to connect it to the solar corona.  In the following we also
use results of previous works on the
solar-stellar connection in X-ray band plus some stellar data already
published in different contexts; our scope is to draw a global
perspective on coronal structure.

The paper is organized as follows: section 2 describes our method to
use the solar corona as a template to study stellar coronae, section 3
presents and discusses some of the related results along with related
works on stellar X-ray emission, in section 4 we draw our conclusions.

\section{The "Sun as an X-ray star" program}

The scope of this program has been to study the X-ray emission of the
whole Sun as an X-ray star, and the contribution of the various coronal
structures to its emission, using Yohkoh/SXT data; we have translated
the solar observations into a format and into a context homogeneous to
those of stellar observations.  Thanks to this approach, we can now
easily perform comparative studies among solar and stellar
observations.

Yohkoh/SXT data have relatively high time resolution, down to a few
seconds, and cover uninterruptedly approximately ten years, from the
maximum of the solar cycle in '91 to the following one. All sorts of
phenomena have been observed, ranging from extremely large flares to
moments of very low emission, during the cycle minimum.  The spatial
resolution (typically $\approx 5 "$) allows us to identify and study,
with high spatial detail, the various regions origin of the X-ray
emission and to discriminate, as well as to isolate, the contribution
of different coronal structures to the coronal X-ray emission.

In brief, from the Yohkoh/SXT images of the Sun at a given time, we
derive a $512 \times 512$ pixels map of coronal plasma temperature and
the corresponding map of emission measure; from them we derive the
distributions of coronal emission measure vs.  temperature at each
observation time; then, summing the optically thin spectrum emitted at
each temperature (obtained with the corresponding emission measure
value and a spectral synthesis code) we obtain the spectrum of the
whole corona; folding it with the known instrumental characteristics
(effective area vs.  photon energy and photon redistribution matrix),
we derive the stellar-like focal plane spectrum of the Sun, as it would
be observed with the X-ray telescope of interest (Orlando et al., 2000
- henceforth Paper I; Peres et al., 2000 - henceforth Paper II). In
this work we will focus on the ROSAT/PSPC data set and will use the
ROSAT standard data analysis also to analyze solar data translated into
stellar format and conditions.

The capability to isolate the various kinds of coronal structures has
allowed us to study their conditions and their contributions to the
coronal spectrum (for the flares, see Reale et al., 2001 - henceforth
Paper III; for the background corona, the active regions and the core of active
regions, see Orlando et al.  2001 - henceforth Paper IV).  The good
time coverage of Yohkoh/SXT observations has been exploited to study
how the emission measure distribution and the X-ray spectrum of the whole
corona changes along the solar cycle, with a regular time sampling and
taking care not to include flare events (Paper II and Paper IV).

\section{The F$_X$ - HR diagrams}

\subsection {The diagrams with stellar data}

\begin{figure}[!ht]
\centerline{\psfig{figure=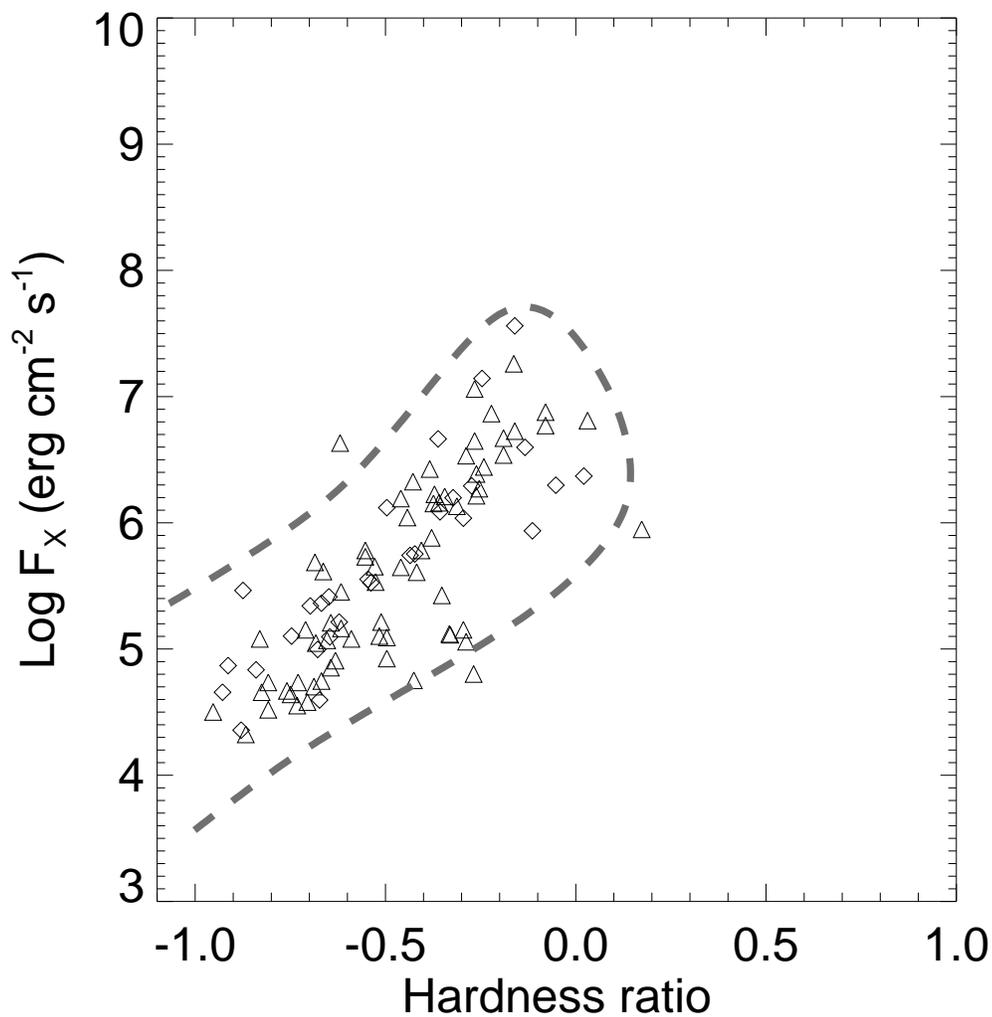,width=15cm,angle=0}}
\vspace{0.3cm}
\caption{$F_X$ vs. HR diagram (see text for details) adapted from
         Schmitt (1997) of a large set of late type stars observed with
         ROSAT/PSPC; diamonds
         identify F and G type stars and triangles K and M stars.
         The dashed area marks the strip covered by late
         type stars in Schmitt (1997) for ease of comparison with the
         following graphs.  We have omitted the stars with HR $\approx
         ~-1$ from Schmitt (1997) because their temperatures are not
         well determined.
	\label{fig1}}
\end{figure}

\begin{figure}[!ht]
\centerline{\psfig{figure=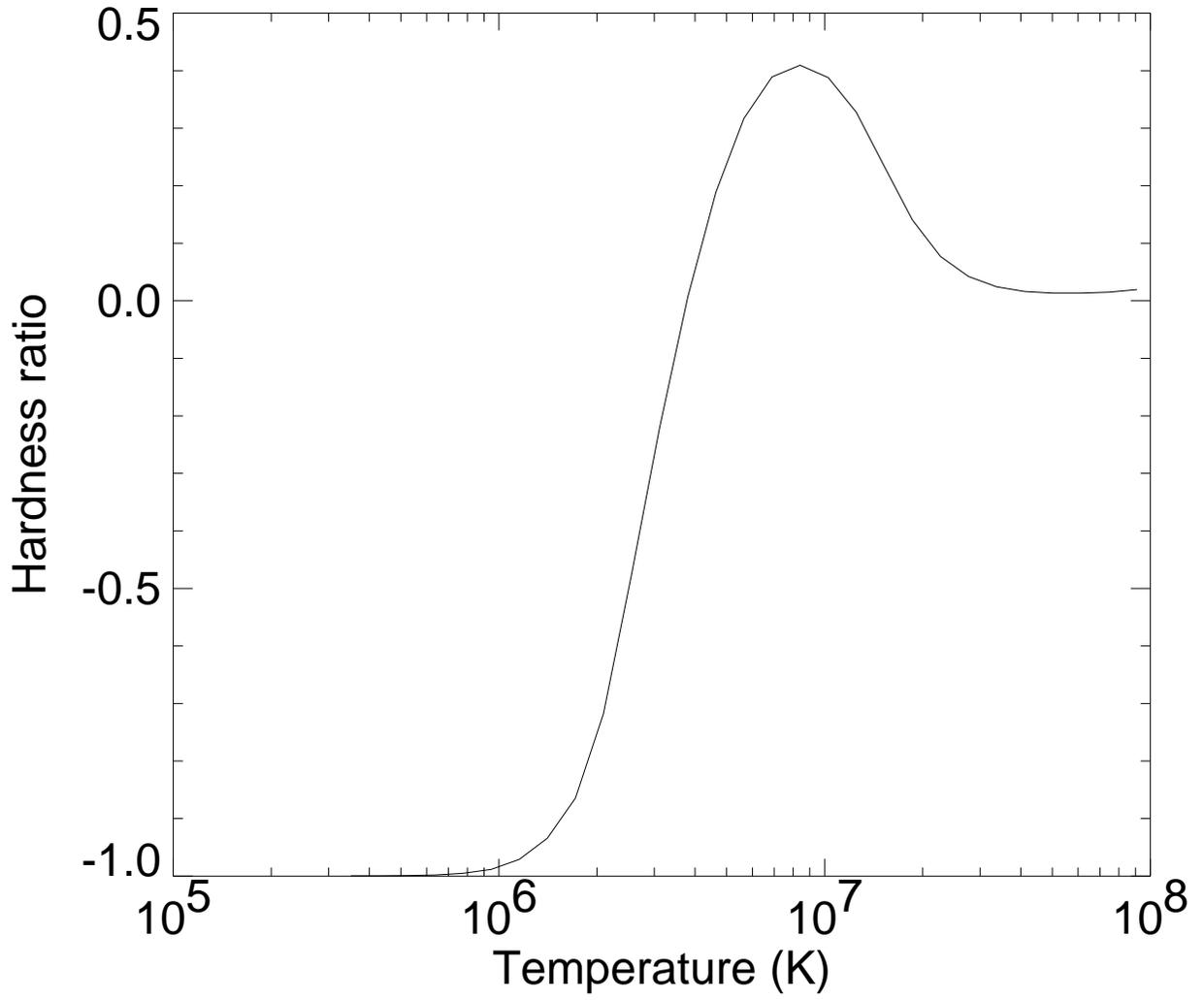,height=15cm,angle=0}}
\vspace{0.3cm}
\caption{The HR (hardness ratio) of the spectrum of a hot single-temperature
         coronal plasma,
         detected with ROSAT/PSPC, vs. the plasma temperature.
	\label{fig2}}
\end{figure}

S97 presented the results of a large set of ROSAT/PSPC observations of
late-type stars, within 13 pc from the Sun, in the form of a diagram
(Fig.  \ref{fig1}) of F$_X$ (the average X-ray surface flux of
the star) vs. HR (the X-ray spectral Hardness Ratio).  The Hardness
Ratio is defined as $HR= ~(H-S)/ (H+S)$ - where H is the total flux in
the ROSAT/PSPC spectral channels between 0.55 and 1.95 keV, and S the
total flux in the channels between 0.13 and 0.40 keV. It is a simple
characterization of the relative weight of the hard and soft part of
the spectrum
\footnote {It is worth
noting that we have omitted, from the sample of S97 stars, those with
$HR \approx -1$, i.e.  $T \approx 10^6$ K since the HR becomes
inadequate (see also Fig.  \ref{fig2}); also small HR errors around
that temperature yield large temperature errors.}.

Both F$_X$ and $HR$ are useful to characterize the strength of coronal
activity, albeit averaged over the whole corona.  Indeed, F$_X$ gives
the coronal radiative losses of the star {\em per unit area} and is
related to the amount of energy delivered into the corona, i.e. the
coronal heating, also
{\em per unit area}; it can discriminate between two stars of the same
X-ray luminosity but of very different radius and surface area: the
smaller star will have a much higher flux than the other, hinting at a
much higher level of plasma activity; F$_X$ is an activity marker
independent of the stellar surface value.  HR is a simple tracer,
albeit a non-linear one, of the coronal average temperature.  It is
another index of plasma activity, independent from F$_X$.

Fig. \ref{fig2} shows the relationship between the plasma temperature
and the HR of the spectrum derived with the Me--Ka--L (Mewe, Lemen,
and van den Oord 1986; Kaastra, J.S. 1992; Mewe, Kaastra and Liedahl,
1995 and references therein) spectral synthesis model.  Note that HR is
little sensitive to temperatures below $10^6$ K; it also saturates and
is multi-valued in the range 0.0 - 0.4; therefore HR is also relatively
insensitive to very hot plasma.  Moreover HR should not go beyond 0.4 at
any plasma temperature.

The S97 results show that late type stars cover a diagonal strip, with
HR ranging from -1.0 to 0.0 and F$_X$ from $10^4$ to $10^8 ~erg
~cm^{-2} ~s^{-1}$ (Fig. \ref{fig1}). The more active a star,
the higher are both its F$_X$ and its HR; the highest HR (temperature)
and F$_X$ values belong to uncommonly active single stars.  The overall
trend suggests that the confined plasma in the corona of a late type
star, on the average, gets brighter as it gets hotter.

\begin{figure}[!ht]
\centerline{\psfig{figure=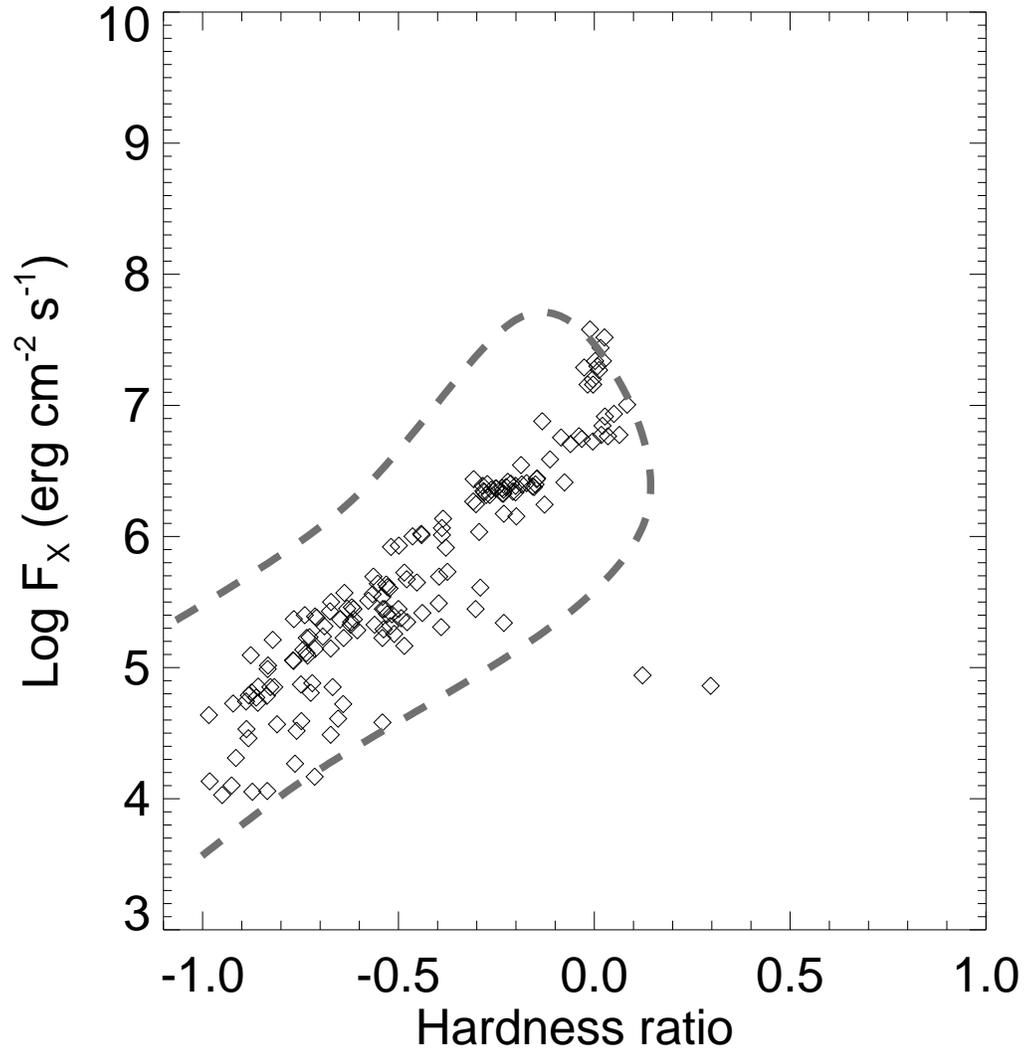,height=15cm,angle=0}}
\vspace{0.3cm}
\caption{Results from Marino et al. (2001) about dF7-dK2 stars, shown
         similarly to Fig. \ref{fig1}.
	\label{fig3}}
\end{figure}

Analogously, Marino et al. (2001) reported the data of 40 late type
stars of spectral type between F7 and K2, within 25 pc from the Sun, in
an analogous F$_X$ vs. HR diagram, shown in Fig.  \ref{fig3}. The
stars of this sample are, on the average, more luminous and active than
those in S97; also, at variance from S97, M and late K stars were not
included.  There were 70 pointed ROSAT/PSPC observations and some of
these stars were observed several times; also some observations, of
sufficient signal to noise ratio, were split to allow time-resolved
analysis during the observation.  Some of these stars moved, in time,
along the same diagonal track where most of the stars are located;
however some stars evolved along a path in this graph, less steep; we
claim, in the following discussion, that this different evolution may
be due to the contribution of flares.  Also Steltzer (2001) generated
preliminary F$_X$ vs. HR diagrams for stars in clusters.

\subsection {The diagrams with solar data}

In Paper II we studied the solar corona at the maximum, at an
intermediate phase and at the minimum of the solar cycle, taking care
not to include flare events; also the whole solar corona during a large
flare event was studied.  Paper III was entirely dedicated to study,
with the same method, the evolution of eight representative flare
events ranging from small to large ones and covering most of the
conditions encountered in flares; each flaring region was studied
separately from the rest of the corona {\em so the surface flux and the
HR were derived just for the flaring region and its effect was not
added to, and diluted within, the whole corona.}

Paper IV studied the solar corona at several moments of the solar
cycle, sampling the cycle better than in Paper II and studying
separately the various classes of coronal structures, classified as
background corona (very quiet regions), active regions and cores of
active regions, according to their surface brightness in Yohkoh/SXT
bands.  The set of whole the data mentioned above yields a good
sampling of the physical conditions in the corona from the background
corona to the strongest flares and span the whole solar cycle.

{\em For the present work
we have taken all the flares in the sample of Paper III because
they are well observed so they can be studied and
characterized accurately; also, since they span a large range within the
flares classification
and the various phases of their evolution are well observed, they
provide a good sampling of the various conditions encountered in
flares, from the largest down to very small events.  Most of the
attention, in the following will be on the physical conditions of
various flare phases. We did not consider micro-flares because they can
hardly be discriminated from small AR variability and their
contribution in the context of average stellar $F_X$ and HR is, in
practice, the same as bright ARs.}

\begin{figure}[!ht]
\centerline{\psfig{figure=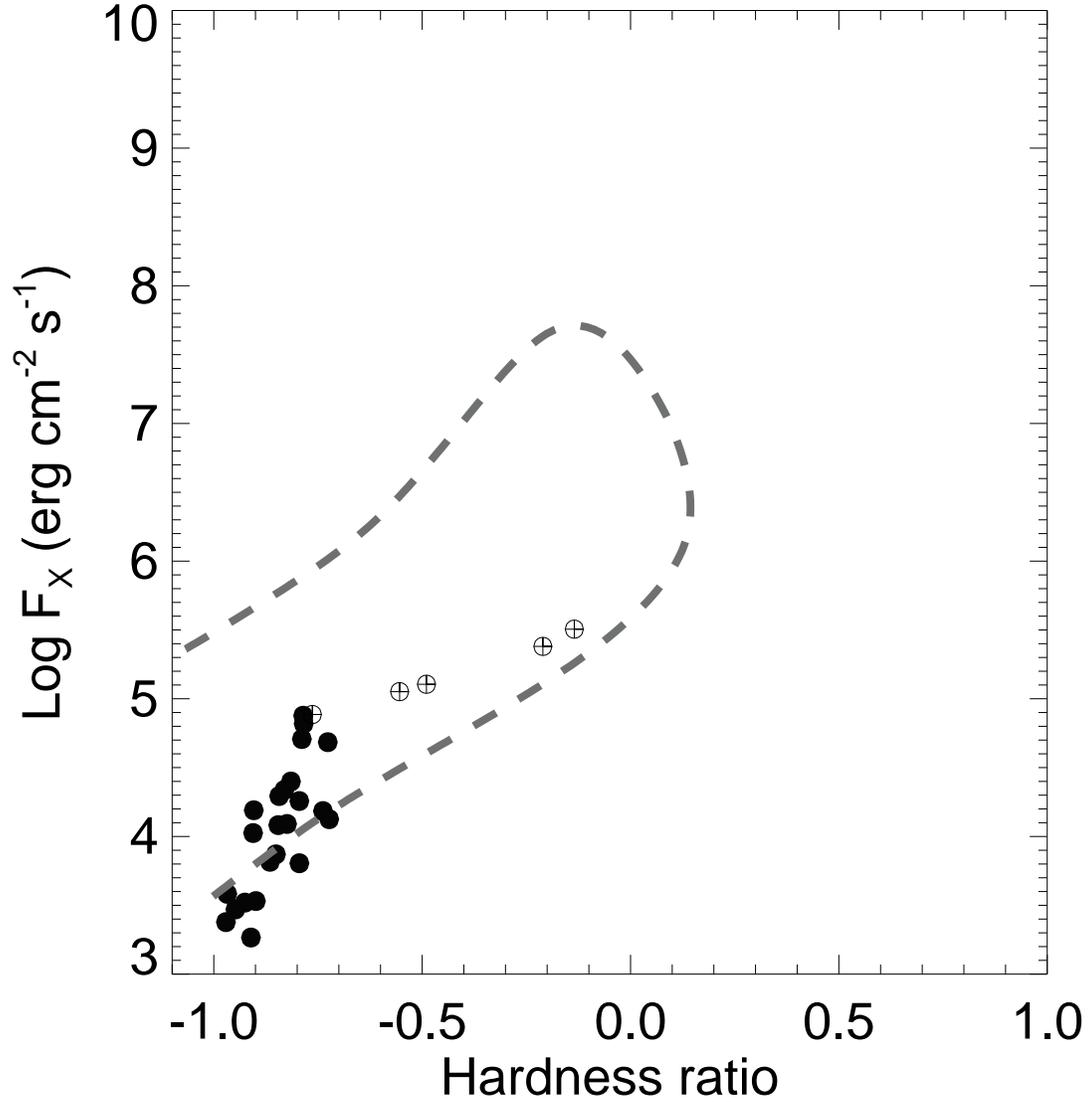,width=15cm,angle=0}}
\vspace{0.3cm}
\caption{F$_X$-HR diagram of the whole solar corona along the solar cycle.
        The data points yield F$_X$ and HR values averaged over the
        whole Sun; filled circled: the solar
        cycle from maximum (highest values) to minimum (lowest ones);
        circles with a cross: the whole solar corona during an X-9 flare
        (i.e. the evolution of the flare plus the steady non-flaring corona).
        Data are from Peres et al. (2000), from Orlando et al.
        (2001), and from Orlando et al. (2004).
	\label{fig4}}
\end{figure}

Fig. \ref{fig4} shows a $F_X$ - HR graph of the whole solar corona
during the cycle plus the evolution of the whole solar corona, near
maximum of the cycle, in the course of a X-9 flare; the graph collects
some results of Paper II, of Paper IV and of Orlando et al. (2002).
The flux of the whole solar corona was derived dividing the total
corona luminosity by the whole solar surface area.

The evolution of the whole solar corona along the solar cycle, apart
from
{\bf moderate to large}
flares, spans the lower part of the diagram, the same covered by low
activity stars, i.e. the ones at low F$_X$ and low HR, well below the
levels of most active stars \footnote {Yohkoh/SXT most likely
underestimates the emission of plasma with a temperature around $10^6$
K or lower, as well as the relevant emission measure, while the same
plasma would have significant emission in the ROSAT/PSPC band (see
Judge, Solomon, Ayres, 2003). The solar data points relative to background
corona and/or the very quiet Sun at cycle minimum may have to be
shifted at higher flux; this effect is not crucial, however, in the
context of this paper.}. {\bf Solar flares span a quite large range, from
very faint, barely detectable, to very large ones;
the flares included in our work are those that at least triggered the
Yohkoh flare mode; in this respect they can be considered to be representative
of flares from
moderate to large.}
If a {\bf moderate to large} flare happens at maximum of the cycle, it
makes the luminosity (and the flux) of the solar corona change at most
by a factor $\approx 2$, during large flares. However, the
corresponding increase of HR can be larger than the changes occurring
along the solar cycle, for the same increase of luminosity.  Analogous
flares
occurring during the solar minimum carry a similar change in HR but a
much larger change of luminosity and flux.  Thus, under the effect of
{\bf significant}
flares, the whole Sun moves rightward in the graph, on a path less
steep than that covered by stars of various activity levels.

The occurrence of large flares may explain why the evolution of some
stars studied by Marino et al. (2001) follows paths, in this diagram,
less steep than the track of all the other stars:  albeit they show no
evidence of large changes of flux, they may undergo a moderate flaring
episode which shows up through the HR change and the different path of
the star in the diagram (see also Orlando et al, 2002, 2004).

\begin{figure}[!ht]
\centerline{\psfig{figure=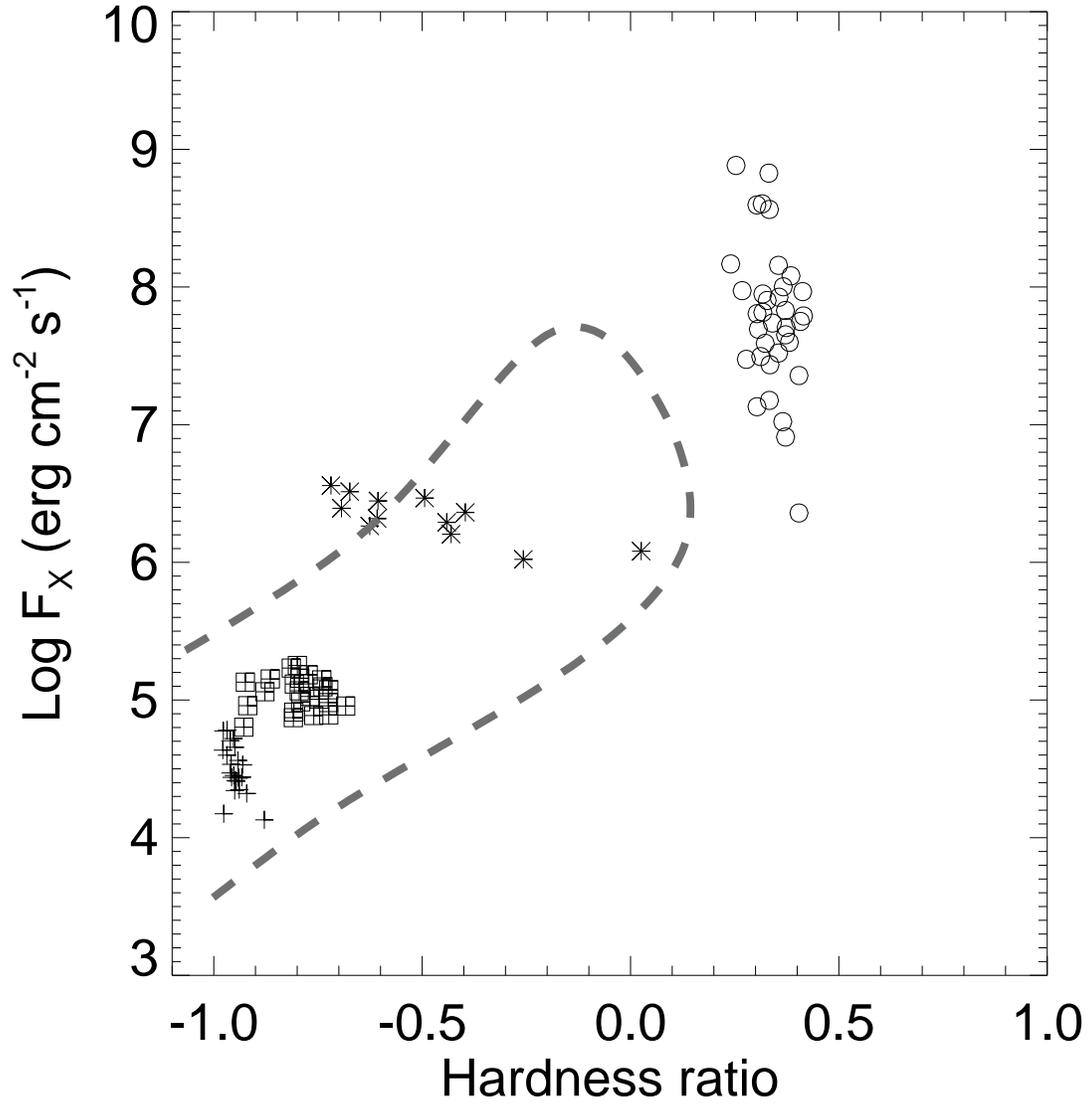,width=15cm,angle=0}}
\vspace{0.3cm}
\caption{
	F$_X$ - HR diagram of various kinds of structures present in
	the solar corona: crosses - background corona; squares - active
	regions; asterisks - cores of active regions; circles - flares.
	Data on flares are from Reale et al. (2001), all the others are
	from Orlando et al.  (2001).
        \label{fig5}}
\end{figure}

Fig. \ref{fig5} shows F$_X$ vs. HR for four specific classes of
structures (i.e. background corona, active regions, cores of active regions
and flares).  It is worth noting that the flux has been derived by
dividing the luminosity of the structure (or of the structures of that
class) by the surface covered in Yohkoh/SXT images.

The interpretation
of these results is useful for stellar physics:  a star entirely covered
with solar-like active regions would be in the same part of the F$_X$-HR
diagram where solar active regions are, and analogously for all the other
kind of structures, since they would yield the same HR (average
temperature) and surface flux values.  Therefore we can foresee how
differently a stellar corona would behave if covered with the various
kinds of solar coronal structures.  {\em More in general, however,} a mixture of two different
classes of structures would yield values intermediate between the two,
the effective value depending on the relative fractions of stellar
surface covered. Interestingly enough, solar structures of different
kinds (with the exception of solar flares) span the same region
of the graph covered by stellar coronae of different activity levels.
{\em Quite likely the stellar surface is not entirely covered with X-ray
emitting structures and different classes of structures
(according to our classification) do coexist on the star, covering
different fractions of its surface. While places at the extremes of the
graph can most likely be explained with only one kind of structure, a locus
between two classes of structures 
may be explained with an appropriate (non-linear) combination of the two.}
{\bf It is worth noting that, according to the analogy we are drawing here, a
star equivalent to a structure, e.g. an active region, would be located
in the same place in the diagram if it were fully covered by this type
of structure; similarly a star intermediate between two kinds of
structure should have its surface close to 100\% covered with the
two kinds of structures. In the solar case we know from Yohkoh/SXT observations
that active regions, their cores and flares (when they occur) take a
small fraction of the surface of the Sun, the rest taken by the very faint
background corona (as shown by very long exposure images). S97 has shown
that very faint coronae have a surface flux comparable to that of very
faint solar background corona. We can, therefore, 
imagine that also on active stars
a faint background corona covers the surface 
not covered by active structures.}

On the basis of the above interpretation, then, we see that the coronae
of stars with extremely low surface flux and hardness ratio could be
just a large solar coronal hole as, on the other hand, S97 suggested.
Continuing this exercise, by increasing the fraction of stellar surface
covered with active regions, or with the cores of active regions,{\em
or a combination thereof,} the corona spans the strip covered by
coronae of late-type-stars from low to intermediate activity.

Therefore we can draw a {\em hypothesis according to} which the levels
of stellar coronal X-ray emission are due to different fractions of the
stellar surface covered with active regions or even cores of young
active regions. The fraction of more active components increases going
towards coronae of higher $F_X$ and higher HR.  With one notable
exception: the few, very-high-$F_X$ ($> ~10^{6.5} ~erg~cm^{-2}~s^{-1}$)
and very-high-HR ($>~~-0.3$), late type stars. The {\it average}
characteristics of these few stars are beyond the {\it peak} values of
the solar coronal active regions cores.  Yet they are well below
{\bf conditions of solar flares of interest here}.  The most active late type
stars can, therefore, be explained as being partly covered with active
regions, their cores and with flares or flare-like (i.e. very hot and
dense) structures {\it at all times}.  Very active stars may then be
continuously subject to several uncorrelated flares whose light curves
overlap randomly, canceling most of the evidence of variability.  Some
authors have already proposed the presence of continuous flaring
activity in stellar coronae (Wood et al. 1997; Drake et al. 2000;
Guedel et al. 2002, and references therein).

Flares can
contribute to coronae of low activity on the strip as long as the
flares are small; indeed large flares in low activity stars are very
evident and make their $F_X$ - HR point move along a path much less
inclined than the strip, bringing them out of the strip itself.
There appears to be no match in the solar corona for those
super-hot and super-luminous flares detected on very active stars and
on RS CVns with the Beppo-SAX and ASCA satellites (Franciosini et al.,
2001; Maggio et al.  2000; Favata et al. 2000; Favata and Schmitt,
1999; Schmitt and Favata, 1999).

\subsection{Global scaling relations}

The $F_X$ - HR diagram provides information on the average features of
the coronal loops forming the structures.  The amount of energy emitted
from the coronal part of any loop is approximately one half of the
entire heating of the loops per unit cross-sectional area
(approximately equal to the part of the solar surface covered by the
loops footpoints), the other half being emitted from the transition
region (Vesecky et al., 1979).   Therefore, considering that $F_X$ is
proportional to the heating

$F_X = ~K~F_{tot}= ~K~E_H~L$

and, 
using the scaling laws by Rosner, Tucker and Vaiana (1978), in the
following RTV78, i.e.

\begin{equation}
E_H= ~10^5~ p^{7/6}~ L^{-5/6} \label{eq:RTVEH}
\end{equation}
and
\begin{equation}
T= ~1.4 \times 10^3 ~(p~L)^{1/3} \label{eq:RTVT}
\end{equation}

\noindent
we find $F_X = ~ K ^ \prime ~p^{7/6}~L^{1/6} =~ K ^ \prime ~p~(pL)^{1/6}= ~K ^{\prime \prime}~p~T^{1/2}$

\noindent
where $K$, $K ^ \prime $ and $K ^{\prime \prime}$ are proportionality
factors, $F_{tot}$ is the total radiated flux, $E_H$ is the heating per
unit volume averaged along field lines, p is the coronal plasma
pressure, T is the coronal loop {\bf maximum} temperature and L is the half-length of
the magnetic loop.  Then, given the very weak dependence on T, $F_X$ is
very close to being linearly related to the coronal pressure.  A closer
look to the above derivation shows that, on one hand, the linear
relation between $F_X$ and $F_{tot}$ is appropriate only for a large
part (but not all) of the temperature range, since one expects that for
very high T (above a few $10^7$ K) or very low T (below $10^6$ K)
different fractions of $F_{tot}$ fall outside the ROSAT band.
Nonetheless the main point of the above exercise is to show that $F_X$
variations are largely dominated by pressure variations.  Of course, HR
changes are related to temperature changes.

\begin{figure}[!ht]
\centerline{\psfig{figure=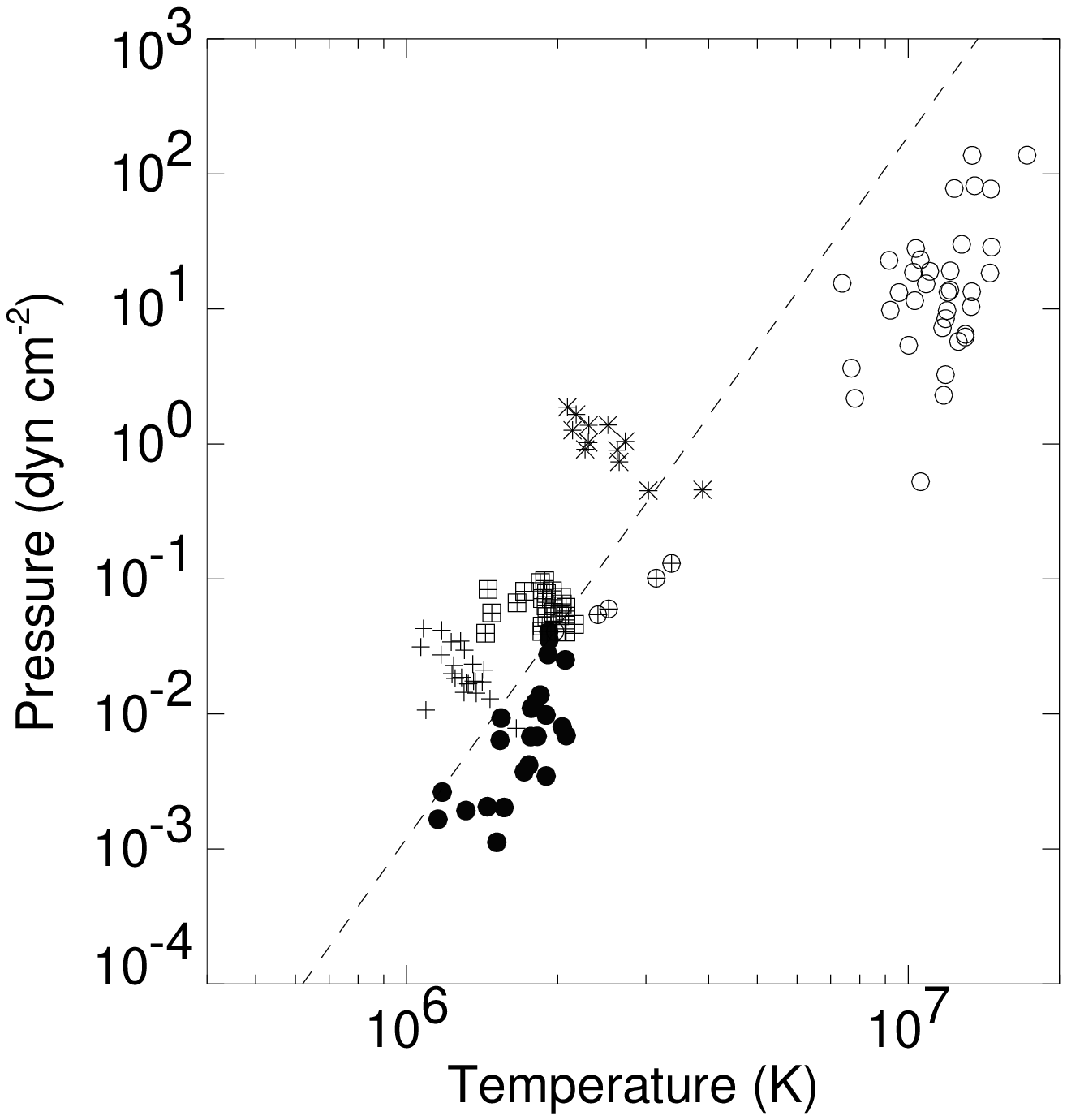,height=7cm,angle=0}}
\centerline{\psfig{figure=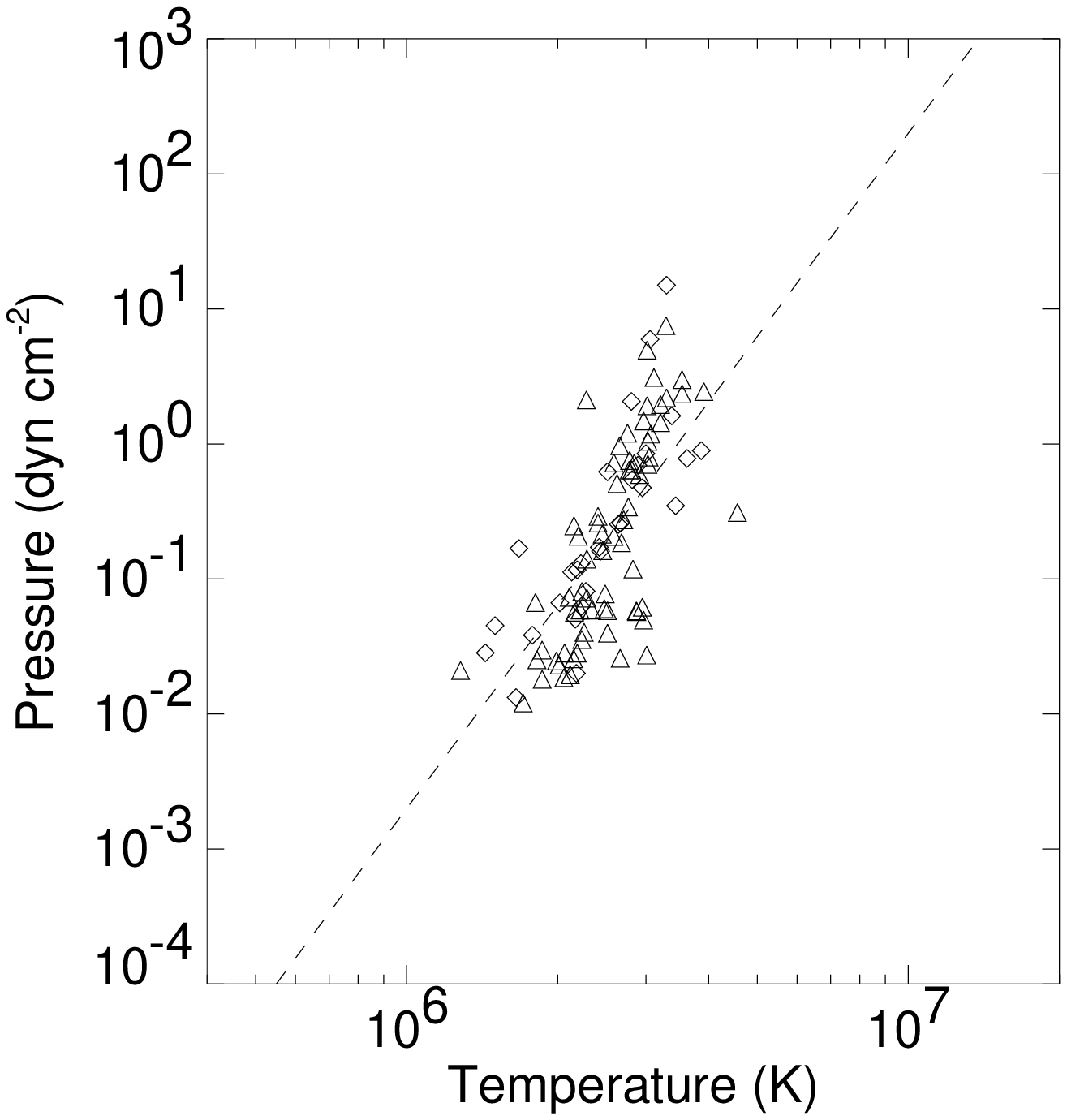,height=7cm,angle=0}}
\centerline{\psfig{figure=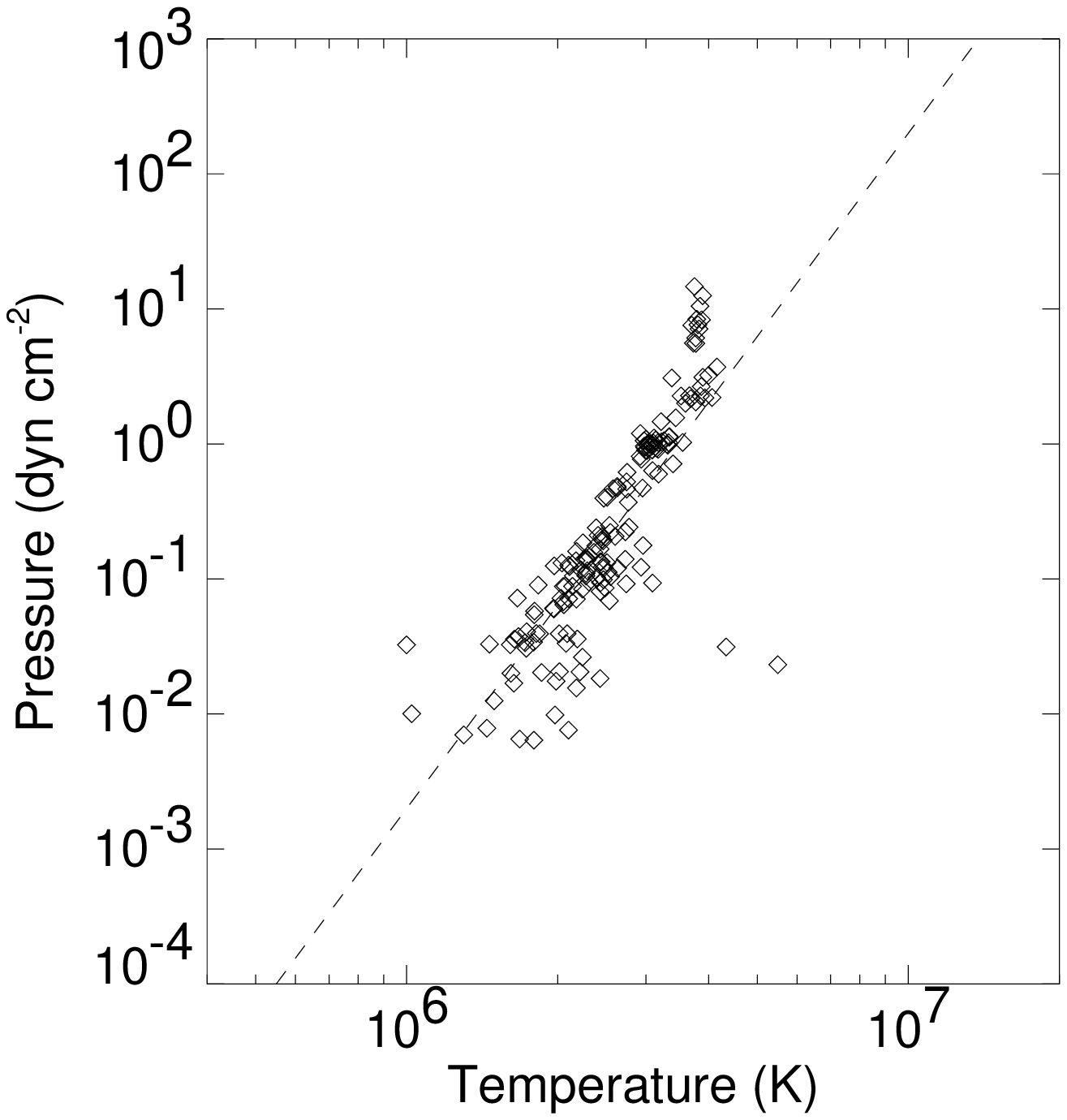,height=7cm,angle=0}}
\vspace{0.3cm}
\caption{Pressure-temperature diagram for the stellar and the solar
	 data shown in previous figures. Top panel: whole solar corona
	 during the cycle, shown in Fig. \ref{fig4} plus the various
	 structures shown in Fig.  \ref{fig5}.  Middle panel:
	 stellar data from Schmitt (1997).  Bottom panel: stellar data
	 from Marino et al. (2001).  Symbols are the same as in previous
	 figures. Dashed line: the $p= 1.2 \times 10^{-3} ~{T_6}^{5.2}$
	 fitting to the data in Marino et al. (2001).
	\label{fig6}} \end{figure}

Using the simple relation derived above and the HR vs. T relation in
Fig. \ref{fig2}, it is straightforward to translate the $F_X$ vs. HR
diagram into an "average pressure" vs. "average temperature" diagram,
deriving the temperature from the single-valued part ($T<~4 \times
10^6$ K) of the hardness ratio and then the pressure as $p= F_X/(
K^{\prime \prime}~T^{0.5})$.  Fig.  \ref{fig6} shows the
pressure-temperature graphs for solar structures, for the whole solar
corona, for the F and G stars in S97, and for the G stars (from F7 to
K2) in Marino et al. (2001)

The data set by Marino et al. (2001) is the most homogeneous one of all
those reported here.  We have fitted the relevant data, finding the relation
\begin{equation}
p =~ 1.2 \times 10^{-3} ~ {T_6}^{5.2} \label{eq:pT}
\end{equation}
shown as a dashed line in all the graphs in Fig.  \ref{fig6} ($T_6$ is
the temperature in units of $10^6$ K);  more precisely the power index
found from the fitting is $5.2 \pm 0.3$.  It is quite interesting that
all the stars, the solar corona during the cycle and the various coronal
structures follow this law, which shows a rather steep dependence of the
confined plasma pressure on the plasma temperature, probably governed
by the heating mechanism(s) due to the magnetic field dissipation.

{\em Flares appear to depart, to some extent, from such a law,
more specifically there appear to be a flattening of the pressure -
temperature relation with slightly lower pressure than that predicted by
the above scaling law.} 

The diagonal strip covered by the solar corona at various phases, and by
solar coronal structures is a coarse but firm relation between average
pressure and temperature of the loops at different evolutionary stages
of the solar structures they form, from emergence to decay.  There is an
analogous, striking, relation for stellar coronae at different activity
levels.

Steady loop models cannot be applied to the rise phase of flares.  However
they can be applied, with caveats, to flares during the slow evolution
of the decay phase. Because of this, the pressure values corresponding
to solar flares are less precise than all the others and have to be
taken with care. Indeed the large spread of solar-flares-related points
suggests the limits of using RTV78 scaling laws on flares.

We also note that the pressure values of the solar structures appear
slightly lower than {\em the values predicted according to eq. (\ref{eq:pT}}); we should consider, however, that the
surface considered, in the Yohkoh/SXT image, is invariably not uniformly
filled with loops. Therefore the local X-ray flux (and the average
pressure) is higher than what we measure, even for the solar structures.
Determining in detail the area really filled with loops, from each
Yohkoh/SXT image, is a formidable task.  We note in passing that this
choice simply skips on the problem of the surface filling factor by
coronal structures, probably present also below the angular
resolution (see however Appendix A).

\begin{figure}[!ht]
\centerline{\psfig{figure=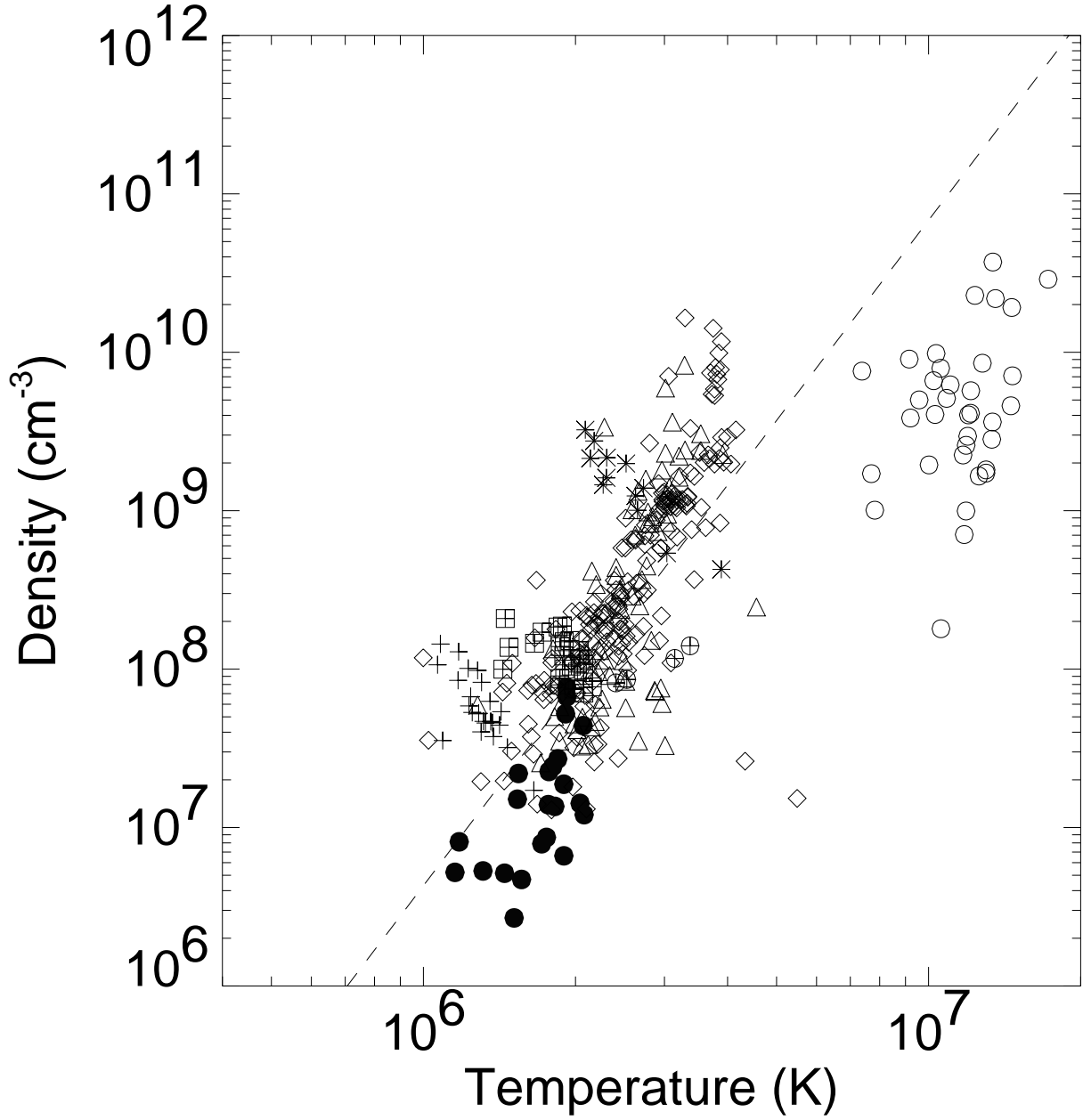,width=18cm,angle=0}}
\vspace{0.3cm}
\caption{Density-temperature diagram of the stellar and of the solar data
	 presented in Fig. \ref{fig6}.
	 The dashed
	 line is the $n =~4.3 \times 10^6 ~{T_6}^{4.2}$ power law derived
         from the $p
	 \propto ~T^{5.2}$ power law and from the ideal gas law for the
	 plasma. Symbols are the same as in the previous figure.
	\label{fig7}}
\end{figure}

An equivalent, albeit slightly different, form of the same relation
is shown in Fig.  \ref{fig7} where we show the average density vs.
temperature, derived with the ideal gas law for a fully ionized hydrogen
plasma

$p = ~ 2~n~k_B~T$ 

where $n$ is the Hydrogen number density and $k_B$ is the Boltzmann
constant.  In this figure we have put together all the solar and
the stellar data;
obviously we find
\begin{equation} n = ~4.3 \times 10^6~ {T_6}^{4.2} \end{equation}

\begin{figure}[!ht]
\centerline{\psfig{figure=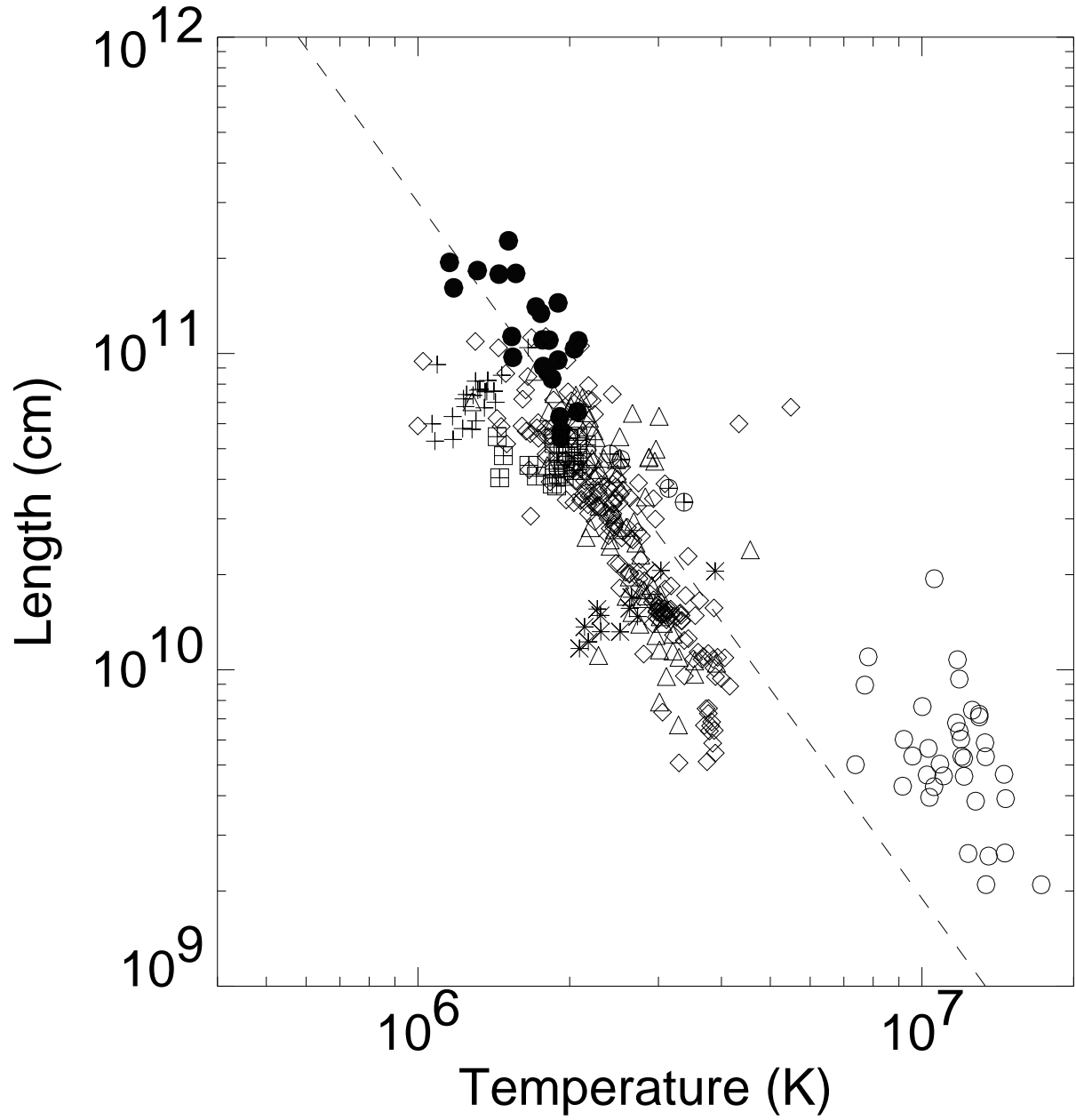,width=18cm,angle=0}}
\vspace{0.3cm}
\caption{Characteristic structures length vs. plasma temperature for the
         data in Fig. \ref{fig6} and in Fig. \ref{fig7}, derived with
         the RTV78 scaling laws.  The dashed line marks the $L= ~ 3.0
         \times 10^{11} ~{T_6}^{-2.2}$ power law. Symbols are the same as
         previous figures.
	\label{fig8}}
\end{figure}

Using the RTV78 scaling law \ref{eq:RTVT}
and substituting there the p-T relation found above,
it is straightforward to translate these
graphs and laws into
the analogous ones relating the 
characteristic structure
length (L) vs. temperature (T). 
Fig. \ref{fig8} shows a graph for the same data as above;
the L-T scaling is
\begin{equation} L= ~3.0 \times 10^{11} ~{T_6}^{-2.2}.  \label{eq:LT}
\end{equation}

\begin{figure}[!ht]
\centerline{\psfig{figure=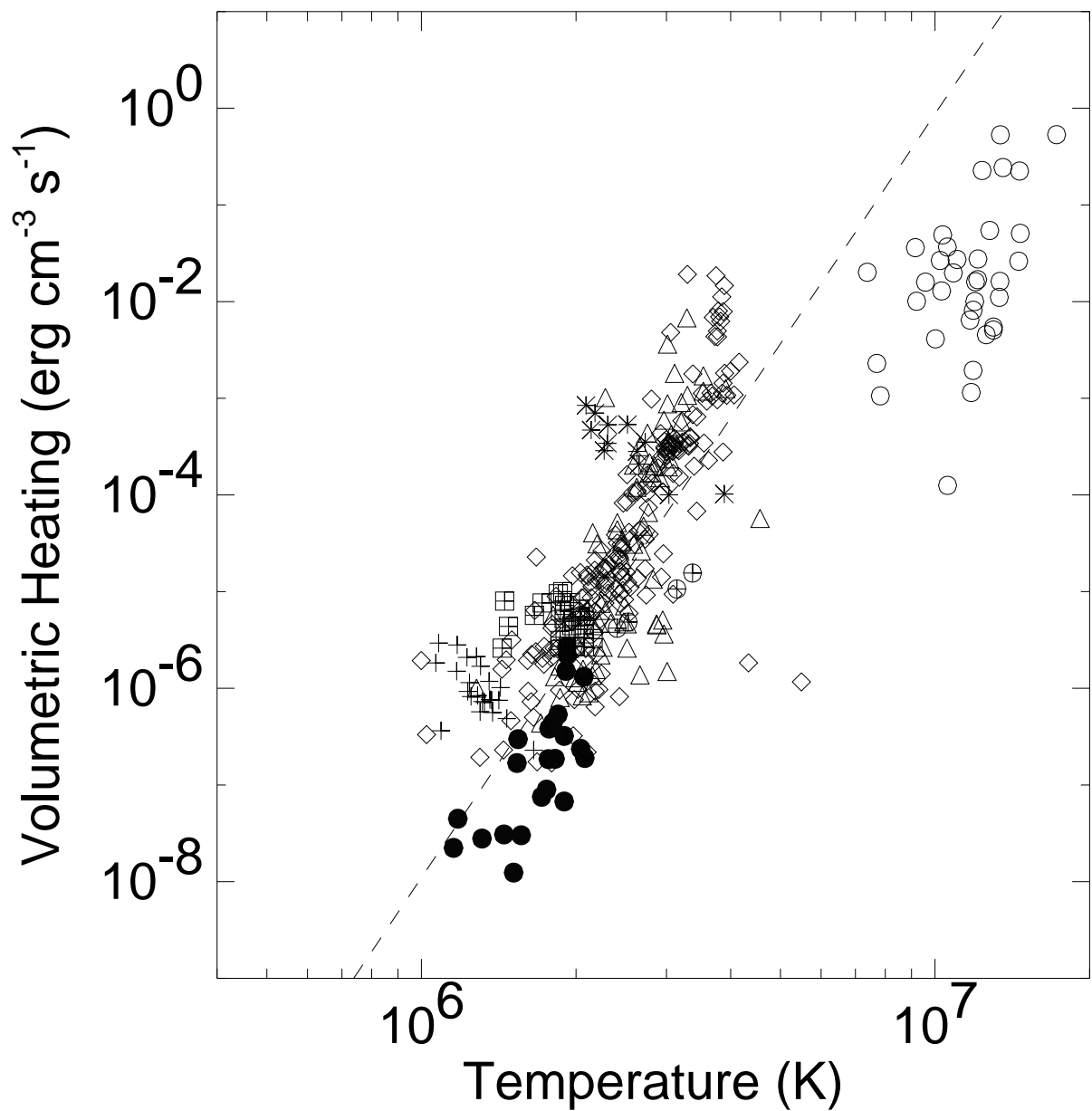,width=18cm,angle=0}}
\vspace{0.3cm}
\caption{Average volumetric heating ($E_H$) vs. temperature for the data
         in Fig.  \ref{fig6} and in Fig. \ref{fig7}, derived with the
         RTV78 scaling laws.  The dashed line marks the $E_H= ~ 1.1
         \times 10^{-8} ~{T_6}^{7.9}$ power law.  Symbols are the same
         as in previous figures.
	\label{fig9}}
\end{figure}

Analogously, substituting the relations \ref{eq:pT} and \ref{eq:LT}
into the scaling law \ref{eq:RTVEH}
we obtain the relation
\begin{equation} E_H= ~1.1 \times 10^{-8} ~{T_6}^{7.9} \label{eq:EHT}
\end{equation}
shown as a dashed line in Fig. \ref{fig9} which also shows the same
solar and stellar data in a $E_H$ 
vs. T graph.

{\em Flares appear to depart from all the above scaling laws and, in
particular, to yield a lower volumetric heating than
predicted by the above scaling law, as an implication of the departure
from the pressure - temperature scaling law. This may just be an
implication of the non-stationarity of the relevant X-ray emitting
plasma. However, since many of such data are related to the decay phase,
i.e. when the loops are in quasi-steady conditions and RTV scaling laws
should be applicable, we should leave open the possibility that there is
indeed a difference in the dissipation mechanisms involved in coronal
steady loops and flares.}

Since dissipation mechanisms tend to connect volumetric heating with the
characteristic size, we are interested to derive a volumetric-heating vs.
structure-length law. Combining the above scaling laws we find
\begin{equation}
E_H= ~1.7 \times 10^{33}~~L^{-3.6}, \label{eq:EHL}
\end{equation}
also probably dictated by the dispersal and dissipation mechanisms of
the magnetic field.

All the above derivation does not take into account the fraction of the
stellar surface covered by the X-ray emitting region, i.e. the surface
filling factor, for clarity's sake.  Appendix A contains the
generalization of the derivation, considering the surface filling
factor.

\section{Discussion}

\subsection{Stellar coronae vs. the solar corona and its building blocks}

We have described the coronae of late-type stars of different activity
levels as covered by correspondingly different fractions of structures
similar to those present in the solar corona.  We have classified
and grouped these structures as background corona, active regions, active
regions cores and flares, in order of increasing surface X-ray flux and
hardness ratio.  Indeed, we find the same correlation of $F_X$ vs. HR
for coronae of late-type stars and for coronal structures present on
the Sun. The solar coronal structures, from background corona to large flares,
fall in the same region of the $F_X$ and HR diagram where the late-type
stars coronae, from very inactive to extremely active, are located.

The occurrence of a large flare on a star of medium {\em or low}
activity level should make the star's position move, in this diagram,
toward the right (i.e. with a significant increase of HR) with
relatively moderate increase of $F_X$ (Paper II; Orlando et al., 2002).
{\em Thus significant flares in these stars should be very prominent for
their variability but also because they should make the star move
outside the strip covered by coronae.}

The most active late type stars appear to have both $F_X$ and HR
significantly higher, even than the active region cores, but one can
still explain them assuming a significant presence, at any time, of
flares or of flare-like structures. Although one cannot exclude that
even moderate activity stars may have a contribution due to flares,
most of their behavior can be easily explained with active regions.
Flares, in analogy with the solar case, should lead to relatively fast
evolution along more horizontal paths slightly departing from the main
region of active stars. Many flares may go undetected for a variety of
reasons, most likely for the very moderate increase of stellar flux or
for the low signal to noise of the observation; undetected small flares
may however lead to a spread along the HR direction of the locus of the
stars in the $F_X$-HR diagram. The diagram in Marino et al. (2001),
containing M stars, indeed shows that the M stars are more spread along
the HR direction than the G stars; it is quite fitting that M stars are
more variable than G ones on short time scales (Marino et al. 2002).

Only the very late evolutionary phases of active region cores, going
toward rather low HR but high surface flux appear to deviate a little
from this scenario. However the fraction of surface covered by these
very old cores appear to be rather small and so, their emission may be
irrelevant with respect to that of the whole stellar corona, as it
happens in the solar corona.

\subsection{The power laws and a unified view of the corona}

The X-ray observations made with Skylab, thanks to their several months
coverage, allowed for the first time to study the long term evolution of
the X-ray corona (for a review, see Vaiana and Rosner, 1978). It became
clear that the magnetically confined coronal plasma evolves from the
compact, hot, bright and high pressure loops of emerging active regions
to progressively longer, cooler, dimmer and relatively low-pressure
loops, first of active regions and then of the background corona.  In this
context, therefore, one can also consider the $F_X$ vs. HR diagram for
solar coronal structures as an evolutionary path. The structures move
along the region covered by the various structures starting as very high
$F_X$, pressure and temperature and small L and, as they evolve, become
of progressively lower $F_X$, pressure, and temperature and get longer.

The various structures of confined plasma, when reported in the
pressure-temperature diagram of Fig. \ref{fig6} or, equivalently,
in the others, in Figs. \ref{fig7}, \ref{fig8} and \ref{fig9}, are all
located along a particular track, probably resulting from the physical
effects governing the dispersion and dissipation of the magnetic field
as it evolves from emergence to total dispersion into very background
corona.
Interestingly enough, this track characterizes both individual structures
in the solar corona and the solar corona as a whole at different phases
of the solar cycle. The global characteristics are the results of a sort
of averaging the features of all the structures present in the corona,
with a weight due to the surface coverage and to the relative brightness.
As a consequence, the average conditions of the solar corona at minimum
of the cycle are closer to those of the background corona, even if some
active regions may be present, and those of the corona at maximum are
closer to active region conditions.

The track also describes the global features of the coronae of solar-type
stars.  While in the solar case we are able to resolve the entire set of
structures at the various evolutionary stages, each star instead just
yields one data point in such a diagram, which yields an average over
all the structures present in corona. The coronal configuration of the
solar-type stars, including the Sun, results from the competing effects
of emergence of magnetic field (generated by the dynamo mechanisms inside
the star) and the magnetic field dispersal and dissipation in the outer
stellar layers; as a consequence, the average values corresponding to each
point in the diagram are determined by these competing physical effects.
The higher the emergence rate, the more the point characterizing the
stellar corona moves towards the high-pressure-high-temperature part of
the graph, as we observe on the Sun (Papers II and IV).

In general, stellar coronae should have a complex coronal composition,
according to this scenario, with different fractions of their surface
covered with compact, hot and bright coronal loops and possibly with
interactions among magnetic fields of different AR (Drake et al., 2000).
A more active star has a higher number of continuously emerging AR with
very effective magnetic field amplification and dissipation mechanisms.

We also claim that each star should have a distribution of structures
of confined plasma obeying this sort of universal pressure-temperature
relationship with loop plasma pressure higher for higher average
loop temperature.

We note that flares appear to behave differently, as for the pressure - 
temperature relation and the related heating - temperature scaling law.
As mentioned, this may depend on the non-stationarity of the relevant
X-ray emitting structures but since some of the observations pertain to
the decay flares phase when the departure from stationarity is small, we
should leave open the possibility that heating mechanisms in flares are
indeed different from those of steady structures.

{\em But, if each of the structures evolve similarly and are equivalent
to the solar ones (as we assume) and the various coronae are just the
sum of the same structures, one should detect all the coronae just as a
time-average of identical coronal structures, thus all with the same
characteristics, independently of the number of structures present. So
why a higher emergence rate should lead to higher average surface flux
and temperature?

On one hand, a higher emergence rate of magnetic structures brings more
ARs and so more X-ray emitting regions to the surface per unit time,
leading to higher luminosity and average X-ray flux of the star.

As for the temperature and HR, the emission measure of an AR and of its
core are significantly larger around their emergence (Orlando et al.,
2004); also flares in an AR are more frequent after emergence.
The unresolved corona approximate the time-average of the
structures only if the number of structures of various age visible at
any time is so large to sample all the evolutionary phases, from the AR
emergence to the "final decay", and including flares. We doubt that a
low-activity star has enough ARs to yield a reasonable "average
behavior of structures".  Flares make the "poor sampling" effect even worse.

Typically a "low activity" star is caught in a non-flaring phase and,
if flares occur, they are easily identified and 
typically analyzed separately from the rest of the observation.
Instead, when we observe a very active star, probably one (or more) of
its ARs is in some phase of a flare.  Indeed, in order to explain the
most active stars we need something brighter and hotter than just
active regions, most likely flares, at any time (Drake et al., 2000;
Paper III).

Even on the most active stars, however, we may not get a large enough
number of ARs to sample well the various flare phases, despite we may
observe enough ARs to sample (and average) their non-flaring evolution.
The stars in the highest part of the $F_X$ - HR diagram
appear to vary somehow from one observation to the other
(Micela, private communication). The locus identifying each very active star in
the diagram may wander around a bit, probably because of flares; the
centroid of the stars' loci at any time may yield the average star
behavior and their spread relative to the centroid may yield the
stellar variations, apart from experimental uncertainties and stellar
peculiarities.

The invariable presence of flares probably makes a further big
difference between less active and very active stars.

For completeness we should consider at least two additional effects: 1)
a large number of ARs may subtract substantial amount of surface to the
faint, diffuse, structures; so the corona is slightly biased toward
higher temperatures and brightness with respect to the average of the
structures evolution we would see on the Sun; 2) as mentioned above if
the ARs crowding is substantial, ARs may interact triggering even more
flares than expected from just a sum of solar structures.}

We have shown that we are able to explain the correlation of $F_X$ vs.
HR of late-type-stars coronae present in the data of S97, and all the
scaling laws we have derived, as due to the plasma structures making the
coronae.  Now the focus moves to explaining why {\em X-ray emitting} solar structures follow
that very relationship, a task which probably should be accomplished
within solar physics and should address the mechanisms governing magnetic
field dispersal and dissipation.

\begin{acknowledgements}
We acknowledge useful suggestions by E. Franciosini, A. Maggio, A. Marino,
G. Micela, J. Sanz-Forcada, S. Sciortino and B. Steltzer {\em and by an
anonimous referee who helped to improve the paper}. This work was
partially supported by Agenzia Spaziale Italiana and by Ministero della
Istruzione, Universit\`a e Ricerca.
\end{acknowledgements}

\appendix
\section{Generalization of the global scaling laws to include the
filling factor}

While in the text we have derived the stellar X-ray surface flux by
dividing the stellar X-ray luminosity $L_X$ by the stellar surface $S_*$,
a more correct derivation of the X-ray surface flux at the star
should consider that only the area $A$ on the star may be covered by coronal
structures. Therefore the whole X-ray luminosity should be 
$L_X=~ A~E_H~L$ and therefore the surface flux we have derived should include
a surface filling factor:

$F_X= ~L_X/S_*=~~A~E_H~L/S_*=~~f~E_H~L$

where $f$, the surface filling factor, is the fraction of the stellar
surface covered with coronal structures.

Therefore we would obtain $F_X=~~f~p~T^{1/2}$ and thus, we would have to
relate the $f~p$ product with temperature, rather than pressure alone.
Thus equation \ref{eq:pT} would become \\
\noindent
$f~p =~ 1.2 \times 10^{-3} ~ {T_6}^{5.2}$

\noindent
and we would obtain

\noindent
$f~n \propto T^{4.2}$,

\noindent
$ L/f= ~3.0 \times 10^{11} ~{T_6}^{-2.2}$ instead of equation  \ref{eq:LT}

and 

\noindent
$f^2 ~ E_H= ~ 1.1 \times 10^{-8} ~{T_6}^{7.9}$ instead of equation \ref{eq:EHT}.

\vspace{0.3cm}

\vspace{0.3cm}

\vspace{0.3cm}

\vspace{0.3cm}

\vspace{0.3cm}

\vspace{0.3cm}

\vspace{0.3cm}

\vspace{0.3cm}

\vspace{0.3cm}

\end{document}